\documentclass{article}
\normalsize
\usepackage{multicol}

\usepackage{set space} 

\title{
Analytic eigenenergies of
the
Dirac equation
with finite degrees of freedom
under a 
confining linear potential using basis functions localized in spacetime
}

\author{Kimichika~Fukushima$^{1}$ and Hikaru~Sato$^{2}$\\
\\
$^{1}$ Advanced Reactor Engineering Department,\\
Toshiba Corporation,\\
8, Shinsugita-cho, Isogo-ku, Yokohama 235-8523, Japan\\
$^{2}$ Emeritus, Department of Physics, Hyogo University of Education,\\
Yashiro-cho, Kato-shi, Hyogo 673-1494, Japan}

\date{             }

\begin{document}

\maketitle

Considering the propagation of fields in the spacetime continuum and the well-defined features of fields with finite degrees of freedom, the wave function is expanded in terms of a finite set of basis functions localized in spacetime.
This paper presents
the 
analytic
eigenenergies
derived
for a
confined fundamental
fermion-antifermion pair
under a linear potential
obtained 
from the Wilson loop for the non-Abelian Yang-Mills field.
The 
Hamiltonian matrix of the Dirac equation is analytically diagonalized using basis functions localized in spacetime. 
The squared
lowest eigenenergy (as a function of the relativistic quantum number when the rotational energy is large compared to the composite particle masses) is
proportional to the string tension and the absolute value of the Dirac's relativistic quantum number related to the total angular momentum, consistent with the expectation.
\par
\verb+ +
\par

\begin{multicols}{2}

\section{Introduction}

\begin{spacing}{1.0} 
In the formalism
elaborated in our previous publications \cite{Fuku84,Fuku14}, the
fields are expanded in terms of basis functions localized in spacetime.
A key characteristic
of this formulation based on
finite element theory is that it is possible to apply differentiation unlike
in
the finite difference method. Our method
allows the use of
a basis set of step functions, which is rather different from the formulation by Bender {\it et al.} \cite{BenMS}. In the non-Abelian Yang-Mills case [4-7], the analytic
continuum classical field as a possible vacuum reveals the linear potential, and quantum fluctuations
are
expressed in terms of step functions
exhibiting 
Coulomb potential.
Regarding
the confined bound state of a
fundamental fermion-antifermion pair and
related themes,
one can observe
approaches from field theory and
lattice gauge theory in
prior literature
[8-18].
Although the Dirac equation with a linear potential was investigated analytically by other authors [19-22], the Regge trajectory \cite{ColReg} may not be reproduced systematically in the relativistic scheme.
Concerning a
classical mechanical Hamiltonian that describes the principal properties of the Regge trajectory,
comprising a linear potential and repulsive rotational potential \cite{Miya,Maki}, which is to be described in the part of this paper containing Eqs. (\ref{EMFM})-(\ref{EMTO}) of Subsection 3.1,
the
Hamiltonian
has no basis in the theory of Dirac fields with the potential produced by the Yang-Mills fields. Furthermore, 
a clear answer has not been fully
provided at the quantum level by other theoretical/numerical approaches to the following questions. What is the mechanism of the mass of a pair composed of a fundamental fermion and antifermion? Why can the binding energy (mass) of the paired fermions be large compared with the masses of the composite fermion and antifermion?
The
confined
bound state of composite fermions is
not fully understood theoretically.
Because the Regge trajectory is expressed by the quantum number corresponding to the rotational quantity, it is expected that the Dirac fields, expressed in spherical coordinates with the potential derived directly from the Yang-Mills theory, will provide answer to the aforementioned questions and reveal the meaning of the
classical mechanical Hamiltonian.
\par
This
work
is aimed at obtaining an analytic expression
for the eigenenergies of
a confined fundamental fermion-antifermion pair using
our formalism mentioned above \cite{Fuku84,Fuku14},
which formulates fields of finite degrees of freedom using basis functions localized in the spacetime continuum. The present formalism enables an analytical calculation without using numerical values unlike numerical computer simulations via numerical values.
The action-like total Hamiltonian including a given linear potential, which leads to the Dirac equation in spherical coordinates by variational
calculus, 
is expressed in 
terms
of a basis set of step functions localized in spacetime. The Hamiltonian matrix in the secular equation is diagonalized analytically, and the
lowest eigenvalue is derived as a function of the string tension and
Dirac's relativistic quantum number related to the total angular momentum. The
squared
system energies
for the large rotational energy compared to the constituent particle masses correspond to those in the classical mechanical Hamiltonian case, which
are consistent with
the principal properties of the Regge trajectory \cite{ColReg}. 
We emphasize that (1) the analytical approach shows clearly
that 
the squared system energies are proportional to the string tension and quantum number, and
that they
originate 
from the secular equation structure leading to the absolute value of the quantum number
that includes a sign and
constructs the potential;
(2) the
quantum number,
instead of being a
non-relativistic value,
is
the relativistic integer
described
by
the Dirac equation for a pair of fundamental fermions.
\par
This article is organized as follows.
Section 2 describes the formalism for the Dirac equation using a basis set of step functions localized in spacetime. Section 3 presents
an
analytic expression 
for the
eigenenergies of a confined bound fermion-antifermion pair,
including discussions in Subsection 3.2,
and is
followed by conclusions in Section 4.
\par
\end{spacing} 

\section{Basis equations and theoretical formalism}

\subsection{Dirac equation for the non-Abelian case in spherical coordinates}

To avoid
confusion, we first note that
the 
relativistic $\kappa$ used in this paper has the following relation \cite{Schf} with the Dirac's notation $j_{\rm D}$ \cite{Dira}
\begin{eqnarray}
\kappa=j_{\rm D}=
\left\{\begin{array}{l}
l+1 \hspace{4ex}\mbox{for $j=l+1/2$ }\\
\mbox{ }\\
-l \hspace{6ex}\mbox{for $j=l-1/2$ }
\end{array}\right. ,
\end{eqnarray}
where $l$ and $j$ refer to
the 
quantum numbers for angular and total angular momentums, respectively. For the lowest energy
case
of a hydrogen atom,
$\kappa > 0$.
Denoting the
radial wave functions of fermions in spherical coordinates with radial $r$-axis as
\begin{eqnarray}
\psi_{F}(r)=\frac{F(r)}{r},
\end{eqnarray}
\begin{eqnarray}
\psi_{G}(r)=\frac{G(r)}{r},
\end{eqnarray}
the Dirac equations in natural units become
\begin{eqnarray}
(+m-\frac{\alpha}{r})F-\frac{{\it d} G}{{\it d} r}-\frac{\kappa}{r}G=EF,
\end{eqnarray}
\begin{eqnarray}
(-m-\frac{\alpha}{r})G+\frac{{\it d} F}{{\it d} r}-\frac{\kappa}{r}F=EG,
\end{eqnarray}
where $m$ is the fermion mass and $E$ is the fermion energy.
\par
Considering the forms of $\psi_{F}=F/r$ and $\psi_{G}=G/r$ as well as the two-dimensional integral $\int dr \hspace{0.5ex} r^2$, the total Hamiltonian
that
variationally leads to the above Dirac equations is
\begin{eqnarray}
\nonumber
{\cal H}=\frac{1}{2}\int dr [F(+m-\frac{\alpha}{r})F
-F\frac{{\it d} G}{{\it d} r}-F\frac{\kappa}{r}G
\end{eqnarray}
\begin{eqnarray}
\nonumber
+ G(-m-\frac{\alpha}{r})G+G\frac{{\it d} F}{{\it d} r}-G\frac{\kappa}{r}F
\end{eqnarray}
\begin{eqnarray}
-FEF-GEG].
\end{eqnarray}
We
now
add the linear energy potential derived from the Wilson loop for
a
non-Abelian field. Energy is one component of
the four-vector momentum
and the linear potential constructs
the energy potential combined consistently with the Coulomb potential, the
Coulomb potential $-\alpha/r$
above 
is replaced by $-\alpha/r+\sigma r$. Here, $\alpha=g^2/(4\pi)$ with
a
coupling constant $g$, and $\sigma$
refers to the 
string tension. We then have the total Hamiltonian for the non-Abelian case
\begin{eqnarray}
\label{DPOT} 
\nonumber
{\cal H}=\frac{1}{2}\int dr [F(+m-\frac{\alpha}{r}+\sigma r)F
-F\frac{{\it d} G}{{\it d} r}-F\frac{\kappa}{r}G
\end{eqnarray}
\begin{eqnarray}
\nonumber
                    + G(-m-\frac{\alpha}{r}+\sigma r)G
                    +G\frac{{\it d} F}{{\it d} r}-G\frac{\kappa}{r}F
\end{eqnarray}
\begin{eqnarray}
-FEF-GEG].
\end{eqnarray}
\par

\subsection{Fields expanded in terms of basis functions localized in spacetime}

The present formalism for fields is firstly based on the propagation of fields in the spacetime continuum, and secondly, on the fact that fields are definite in a scheme of finite degrees of freedom. We then expand the fields in terms of basis functions localized in the spacetime continuum, which has a finite number of lattice (grid) points, by realizing the following formulation.
For the considering region in spherical coordinates, we introduce
lattice (grid) points $r_{n}$ $(n=1, 2, ....,N)$ in the radial $r$-axis. The infinitesimal positive $\Delta$ is defined by $\Delta=r_{n+1}-r_{n}$, and $r_{n-1/2}=r_{n}-(1/2)\Delta$ and $r_{n+1/2}=r_{n}+(1/2)\Delta$.
The
basis functions,
which have a superscript
without and with prime, are defined by
\begin{eqnarray}
 \label{EqB1} 
\Omega^{E}_{n}(r)=
\left\{\begin{array}{ll}
1 & \mbox{ for $r_{n-1/2} < r < r_{n+1/2}$}\\
\mbox{ } & \mbox{ } \\
0 & \mbox{ for $r \leq r_{n-1/2}$ or
$r \geq r_{n+1/2}$}
\end{array}\right. ,
\end{eqnarray}
\begin{eqnarray}
 \label{EqB1b} 
\Omega^{E^{\prime
}}_{n}(r)=
\left\{\begin{array}{ll}
1 & \mbox{ for $r_{n-1/2} \leq r \leq r_{n+1/2}$}\\
\mbox{ } & \mbox{ } \\
0 & \mbox{ for $r < r_{n-1/2}$ or
$r > r_{n+1/2}$}
\end{array}\right. ,
\end{eqnarray}
which have the derivatives
\begin{eqnarray}
 \label{EqB20} 
\nonumber
\frac{d\Omega^{E}_{n}(r)}{dr}|_{r=r_{n-1/2}}
=\frac{d\Omega^{E^{\prime}}_{n}(r)}{dr}|_{r=r_{n-1/2}}
\end{eqnarray}
\begin{eqnarray}
=\delta (r - r_{n-1/2}),
\end{eqnarray}
\begin{eqnarray} 
 \label{EqB2} 
\nonumber
\frac{d\Omega^{E}_{n}(r)}{dr}|_{r=r_{n+1/2}}
=\frac{d\Omega^{E^{\prime}}_{n}(r)}{dr}|_{r=r_{n+1/2}}
\end{eqnarray}
\begin{eqnarray}
=-\delta (r - r_{n+1/2}).
\end{eqnarray}
\par
The above basis functions have the following properties
\begin{eqnarray}
\int dr \Omega^{E}_{n}(r) \Omega^{E}_{k}(r)=\delta_{nk}\int dr \Omega^{E}_{n}(r)=\Delta\delta_{nk},
\end{eqnarray}
\begin{eqnarray}
\int dr \Omega^{E}_{n}(r) \Omega^{E^{\prime}}_{k}(r)=\delta_{nk}\int dr \Omega^{E}_{n}(r)=\Delta\delta_{nk},
\end{eqnarray}
\begin{eqnarray}
\int dr \Omega^{E^{\prime}}_{n}(r) \Omega^{E^{\prime}}_{k}(r)=\delta_{nk}\int dr
\Omega^{E^{\prime}}_{n}(r)
=\Delta\delta_{nk},
\end{eqnarray}
where $\delta_{nk}$ is Kronecker's delta, and
considering the overlap between the basis function $\Omega^{E}_{n}$ (or $\Omega^{E^{\prime}}_{n})$ around the point $r_{n}$ and the delta function around $r_{k-1/2}$ and $r_{k+1/2}$, which is the derivative of the basis function around $r_{k}$,
\begin{eqnarray}
\nonumber
\int dr \Omega^{E}_{n}(r) \frac{d\Omega^{E}_{k}(r)}{dr}
=\int dr \Omega^{E}_{n}(r) \frac{d\Omega^{E^{\prime}}_{k}(r)}{dr}
\end{eqnarray}
\begin{eqnarray}
=\int dr \Omega^{E}_{n}(r) [ 
\delta(r-r_{k-1/2 })-\delta(r-r_{k+1/2 })
]=0,
\end{eqnarray}
\begin{eqnarray}
\nonumber
 \int dr \Omega^{E^{\prime}}_{n}(r) \frac{d\Omega^{E         }_{k}(r)}{dr}
=\int dr \Omega^{E^{\prime}}_{n}(r) \frac{d\Omega^{E^{\prime}}_{k}(r)}{dr}
\end{eqnarray}
\begin{eqnarray}
\nonumber
=\int dr \Omega^{E^{\prime}}_{n}(r)[\delta(r-r_{k-1/2})-\delta(r-r_{k+1/2})]
\end{eqnarray}
\begin{eqnarray}
\nonumber
=-\delta_{k,n-1}
+(\delta_{k,n}-\delta_{k,n})
+\delta_{k,n+1}
\end{eqnarray}
\begin{eqnarray}
=\delta_{k,n+1}-\delta_{k,n-1}.
\end{eqnarray}
\par
Using Eqs. (\ref{EqB1}) and (\ref{EqB1b}), we further define
\begin{eqnarray}
 \label{EqeD}
\Omega^{e}_{n}(r)= \frac{1}{2}[\Omega^{E}_{n}(r)+\Omega^{E^{\prime}}_{n}(r)].
\end{eqnarray}
With the help of Eqs. (\ref{EqB1})-(\ref{EqeD}), we get
\begin{eqnarray}
 \label{Eqe1} 
\nonumber
\int dr \Omega^{e}_{n}(r) \Omega^{e}_{k}(r)
\end{eqnarray}
\begin{eqnarray}
\nonumber
=\frac{1}{4}\int dr 
(
\Omega^{E}_{n} \Omega^{E}_{k}
+\Omega^{E^{\prime}}_{n} \Omega^{E}_{k}
+\Omega^{E}_{n} \Omega^{E^{\prime}}_{k}
+\Omega^{E^{\prime}}_{n} \Omega^{E^{\prime}}_{k}
)
\end{eqnarray}
\begin{eqnarray}
=\frac{4}{4}\delta_{nk}\Delta=\Delta\delta_{nk},
\end{eqnarray}
\begin{eqnarray}
 \label{Eqe2} 
\nonumber
\int dr \Omega^{e}_{n}(r) \frac{d\Omega^{e}_{k}(r)}{dr}
\end{eqnarray}
\begin{eqnarray}
\nonumber
=\frac{1}{4} \int dr
(
\Omega^{E}_{n}         \frac{d\Omega^{E         }_{k}}{dr}
+\Omega^{E}_{n}         \frac{d\Omega^{E^{\prime}}_{k}}{dr}
+\Omega^{E^{\prime}}_{n}\frac{d\Omega^{E         }_{k}}{dr}
+\Omega^{E^{\prime}}_{n}\frac{d\Omega^{E^{\prime}}_{k}}{dr}
)
\end{eqnarray}
\begin{eqnarray}
\nonumber
=\frac{1}{4}
[0+0+(\delta_{k,n+1}-\delta_{k,n-1})+(\delta_{k,n+1}-\delta_{k,n-1})]
\end{eqnarray}
\begin{eqnarray}
=\Delta \frac{(\delta_{k,n+1}-\delta_{k,n-1})}{2\Delta}.
\end{eqnarray}
\par
Then, 
the
expansion of
the 
fermion wave functions
\begin{eqnarray}
\label{SolF} 
F(r)=\sum_{n}F_{n}\Omega^{e}_{n}(r),
\end{eqnarray}
\begin{eqnarray}
\label{SolG} 
G(r)=\sum_{n}G_{n}\Omega^{e}_{n}(r),
\end{eqnarray}
uniquely
discretizes
the following terms in the total Hamiltonian
\begin{eqnarray}
\nonumber
\int dr F(r)G(r)
\end{eqnarray}
\begin{eqnarray}
 \label{EqS} 
\nonumber
=\int dr \sum_{n}\sum_{k}
[F_{n}\Omega^{e}_{n}(r) G_{k}\Omega^{e}_{k}(r)]
\end{eqnarray}
\begin{eqnarray}
\nonumber
=\sum_{n}\sum_{k}
[F_{n}G_{k}\int dr \Omega^{e}_{n}(r) \Omega^{e}_{k}(r)]
\end{eqnarray}
\begin{eqnarray}
=\sum_{n}\sum_{k}
[F_{n}G_{k}\Delta\delta_{nk}]
=\Delta\sum_{n} F_{n}G_{n},
\end{eqnarray}
for which we have used Eq. (\ref{Eqe1}).
\par
Similarly, using
Eq. (\ref{Eqe2})
it follows that
\begin{eqnarray}
 \label{EqDeri} 
\nonumber
\int dr F(r)\frac{dG(r)}{dr}
\end{eqnarray}
\begin{eqnarray}
\nonumber
=\int dr \sum_{n}\sum_{k}
[F_{n}\Omega^{e}_{n}(r)
 G_{k}\frac{d\Omega^{e}_{k}(r)}{dr}]
\end{eqnarray}
\begin{eqnarray}
\nonumber
=\sum_{n}\sum_{k}
[F_{n}G_{k}\Delta \frac{(\delta_{k,n+1}-\delta_{k,n-1})}{2\Delta}]
\end{eqnarray}
\begin{eqnarray}
=\Delta\sum_{n} F_{n}\frac{G_{n+1}-G_{n-1}}{2\Delta}.
\end{eqnarray}
\par

\subsection{Matrix form of Dirac equation in terms of basis functions localized in spacetime}

In our scheme, the aforementioned total Hamiltonian in a fermion-confined region is expressed by
\begin{eqnarray}
\nonumber 
{\cal H}=\frac{\Delta}{2}
\end{eqnarray}
\begin{eqnarray}
\nonumber
\times\sum_{n}
\{ [F_{n}(+m-\frac{\alpha}{r_{n}}+\sigma r_{n})F_{n}
\end{eqnarray}
\begin{eqnarray}
\nonumber
-F_{n}\frac{G_{n+1}-G_{n-1}}{2\Delta}-F_{n}\frac{\kappa}{r_{n}}G_{n}]
\end{eqnarray}
\begin{eqnarray}
\nonumber
+[G_{n}(-m-\frac{\alpha}{r_{n}}+\sigma r_{n})G_{n}
\end{eqnarray}
\begin{eqnarray}
\nonumber
+G_{n}\frac{F_{n+1}-F_{n-1}}{2\Delta}-G_{n}\frac{\kappa}{r_{n}}F_{n}]
\end{eqnarray}
\begin{eqnarray}
-F_{n}EF_{n}-G_{n}EG_{n} \}.
\end{eqnarray}
The potentials $\alpha/r_{n}$,
$\sigma r_{n}$
and $\kappa/r_{n}$ are replaced by $\alpha/(n\Delta)$,
$n\sigma \Delta$
and $\kappa/(n\Delta)$, respectively.
Let us impose the normalization condition on $F$ and $G$ and replace the last terms $-F_{n}EF_{n}-G_{n}EG_{n}$ by $-E(F_{n}F_{n}-1/N)-E(G_{n}G_{n}-1/N)$, with $N$ being the number of lattice (grid) points, and $E$ is regarded as a Lagrange multiplier.
\par
By variational
calculus,
$\Delta$ remains in all
the
terms including
those
with
$E$,
and all terms are divided by $\Delta$. We then get the Dirac equation in the matrix form for the $v$-th eigenvector ${\bf x}^{v}$ and
the 
associated eigenenergy $E^{v}$
\begin{eqnarray}
{H}{\bf x}^{v}=E^{v}{\bf x}^{v}.
\end{eqnarray}
A detailed form of the equation
is
\begin{eqnarray}
\left[\begin{array}{cc}
{H}_{A} &{H}_{B} \\
{H}_{C} &{H}_{D} 
\end{array}\right]
\left[\begin{array}{c}
{\bf F}^{v} \\
{\bf G}^{v} 
\end{array}\right]
=E^{v}
\left[\begin{array}{c}
{\bf F}^{v} \\
{\bf G}^{v} 
\end{array}\right],
\end{eqnarray}
where the components of the row vectors ${\bf x}^{v}$, ${\bf F}^{v}$ and ${\bf G}^{v}$ have the relations
\begin{eqnarray}
x^{v}_{i}  =F^{v}_{i},
\end{eqnarray}
\begin{eqnarray}
x^{v}_{N+i}=G^{v}_{i},
\end{eqnarray}
for $ 1 \leq i \leq N $.
It is to be noted that $i$ is not the imaginary unit in complex numbers, but an integer.
The matrix
$H$ 
has the following form
\begin{eqnarray}
\nonumber
H
=
\left[\begin{array}{cc}
{H}_{A} &{H}_{B} \\
{H}_{C} &{H}_{D} 
\end{array}\right]
\end{eqnarray}
\begin{eqnarray}
=
\left[\begin{array}{cccccccc}
 \times &0      &0      &0      &\times &\times &0      &0      \\
 0      &\times &0      &0      &\times &\times &\times &0      \\
 0      &0      &\times &0      &0      &\times &\times &\times \\
 0      &0      &0      &\times &0      &0      &\times &\times \\
 \times &\times &0      &0      &\times &0      &0      &0      \\
 \times &\times &\times &0      &0      &\times &0      &0      \\
 0      &\times &\times &\times &0      &0      &\times &0      \\
 0      &0      &\times &\times &0      &0      &0      &\times \\
\end{array}\right],
\end{eqnarray}
where for $ 1 \leq i, j \leq N $,
\begin{eqnarray}
\hspace{-8ex}
H_{Aij}=H_{ij}\delta_{ij}=(+m-\frac{\alpha}{i\Delta}
+\sigma i\Delta)
\delta_{ij},
\end{eqnarray}
\begin{eqnarray}
H_{Bij}=H_{i,N+j}=
\left\{\begin{array}{l}
\frac{-\kappa}{i\Delta} \hspace{4ex}\mbox{for $j=i$} \\
\mbox{ } \\
\frac{+1}{
2
\Delta
} \hspace{4ex}\mbox{for $j=i-1$} \\
\mbox{ } \\
\frac{-1}{
2
\Delta} \hspace{4ex}\mbox{for $j=i+1$} \\
\mbox{ } \\
0 \hspace{4ex}\mbox{for the others} 
\end{array}\right. , 
\end{eqnarray}
\begin{eqnarray}
H_{Cij}=H_{N+i,j}=
\left\{\begin{array}{l}
\frac{-\kappa}{i\Delta} \hspace{4ex}\mbox{for $j=i$} \\
\mbox{ } \\
\frac{-1}{
2
\Delta} \hspace{4ex}\mbox{for $j=i-1$} \\
\mbox{ } \\
\frac{+1}{
2
\Delta} 
\hspace{4ex}\mbox{for $j=i+1$} \\
\mbox{ } \\
0 \hspace{4ex}\mbox{for the others} 
\end{array}\right. , 
\end{eqnarray}
\begin{eqnarray}
H_{Dij}=
H_{N+i,N+j}
\delta_{ij}=(-m-\frac{\alpha}{i\Delta}
+\sigma i\Delta)
\delta_{ij}.
\end{eqnarray}

\section{Analytic eigenenergies of a fermion under a confining linear potential}

\subsection{Analytic eigenenergies of the Dirac equation with a linear potential}

The quantum eigenenergies are now treated by considering the energies of a classical mechanical Hamiltonian for the principal properties of the Regge trajectory.
We
use the following matrix
$R$
in
the same notation
as
the aforementioned matrix
$H$
\begin{eqnarray}
{R}
=
\left[\begin{array}{cc}
{R}_{A} &{R}_{B} \\
{R}_{C} &{R}_{D} 
\end{array}\right],
\end{eqnarray}
where
\begin{eqnarray}
R_{p_{r},q_{r}}=\frac{1}{C_{\rm N}}[\sin(\frac{2\pi p_{r}q_{r}}{2N+1})],
\end{eqnarray}
for $ 1 \leq p_{r}, q_{r} \leq 2N $, and ${C_{\rm N}}$ is a normalization constant
and large for large $N$.
We then have (emphasizing the matrix elements of
${R}$ 
using parentheses
[\hspace{1ex}])
\begin{eqnarray}
\nonumber
({HR})_{i,q_{r}}
=\frac{1}{C_{\rm N}} \{
\end{eqnarray}
\begin{eqnarray}
\nonumber
H_{ii}[\sin(\frac{2\pi (i)q_{r}}{2N+1})]
\end{eqnarray}
\begin{eqnarray}
\nonumber
+H_{i,N+i-1}[\sin(\frac{2\pi (N+i-1)q_{r}}{2N+1})]
\end{eqnarray}
\begin{eqnarray}
\nonumber
+H_{i,N+i}[\sin(\frac{2\pi (N+i)q_{r}}{2N+1})]
\end{eqnarray}
\begin{eqnarray}
+H_{i,N+i+1}[\sin(\frac{2\pi (N+i+1)q_{r}}{2N+1})] \}.
\end{eqnarray}
The
second and last terms below
``$1/C_{\rm N} \{$''
in the above equation
amount to an expression in term of the matrix component $[{\sin(2\pi (N+i)q_{r}/(2N+1}))]$ of
${R}$
\begin{eqnarray}
\nonumber
H_{i,N+i-1}[\sin(\frac{2\pi (N+i-1)q_{r}}{2N+1})]
\end{eqnarray}
\begin{eqnarray}
\nonumber
+H_{i,N+i+1}[\sin(\frac{2\pi (N+i+1)q_{r}}{2N+1})]
\end{eqnarray}
\begin{eqnarray}
\nonumber
=\frac{1}{
2
\Delta}[\sin(\frac{2\pi (N+i-1)q_{r}}{2N+1})]
\end{eqnarray}
\begin{eqnarray}
\nonumber
+\frac{-1}{
2
\Delta}[\sin(\frac{2\pi (N+i+1)q_{r}}{2N+1})]
\end{eqnarray}
\begin{eqnarray}
\nonumber
=\frac{1}{\Delta}\sin(\frac{-2\pi q_{r}}{2N+1})
\frac{\cos(\frac{2\pi (N+i)q_{r}}{2N+1})}{\sin(\frac{2\pi (N+i)q_{r}}{2N+1})}
\end{eqnarray}
\begin{eqnarray}
\times [\sin(\frac{2\pi (N+i)q_{r}}{2N+1})].
\end{eqnarray}
Thus,
${HR}$ has been
set to
$
{H^{\prime}R}$
with
\begin{eqnarray}
H^{\prime}_{ii}=H_{ii},
\end{eqnarray}
\begin{eqnarray}
H^{\prime}_{i-1,N+i}=0,
\end{eqnarray}
\begin{eqnarray}
\nonumber
H^{\prime}_{i,N+i}=H_{i,N+i}
\end{eqnarray}
\begin{eqnarray}
+\frac{1}{\Delta}\sin(\frac{-2\pi q_{r}}{2N+1})
\frac{\cos(\frac{2\pi (N+i)q_{r}}{2N+1})}{\sin(\frac{2\pi (N+i)q_{r}}{2N+1})},
\end{eqnarray}
\begin{eqnarray}
H^{\prime}_{i+1,N+i}=0,
\end{eqnarray}
where $\kappa$ is contained in
$H_{i,N+i}$.
\par
In a similar way, we
obtain
\begin{eqnarray}
H^{\prime}_{N+i,i-1}=0,
\end{eqnarray}
\begin{eqnarray}
\nonumber
H^{\prime}_{N+i,i}=H_{N+i,i}
\end{eqnarray}
\begin{eqnarray}
+\frac{1}{\Delta}\sin(\frac{2\pi q_{r}}{2N+1})
\frac{\cos(\frac{2\pi (i)q_{r}}{2N+1})}{\sin(\frac{2\pi (i)q_{r}}{2N+1})},
\end{eqnarray}
\begin{eqnarray}
H^{\prime}_{N+i,i+1}=0,
\end{eqnarray}
\begin{eqnarray}
H^{\prime}_{N+i,N+i}=H_{N+i,N+i}.
\end{eqnarray}
Therefore, the matrix
$
{H^{\prime}}$ 
has been reduced to
\begin{eqnarray}
{H^{\prime}}
=
\left[\begin{array}{cccccccc}
 \times &0      &0      &0      &\times &0      &0      &0      \\
 0      &\times &0      &0      &0      &\times &0      &0      \\
 0      &0      &\times &0      &0      &0      &\times &0      \\
 0      &0      &0      &\times &0      &0      &0      &\times \\
 \times &0      &0      &0      &\times &0      &0      &0      \\
 0      &\times &0      &0      &0      &\times &0      &0      \\
 0      &0      &\times &0      &0      &0      &\times &0      \\
 0      &0      &0      &\times &0      &0      &0      &\times \\
\end{array}\right].
\end{eqnarray}
Hamiltonian is Hermitian and ${R}^{\rm t}={R}$, where ${R}^{\rm t}$ is the transpose of ${R}$. The result for the above matrix ${H}^{\prime}$ essentially implies that ${HR}={H}^{\prime}{R}={R}^{\rm t}{H}^{\rm t}={R^{\rm t}}{H}^{\prime{\rm t}}$. From ${H}^{\prime}={R}({R}^{\rm t}{H}^{\prime {\rm t}})$ and ${R^{\rm t}}{H}^{\prime{\rm t}}={HR}$, we get ${H}^{\prime}={R}({HR})={R}^{\rm t}{HR}$, which
implies
that this process is a unitary transformation.
\par
A set of four elements $H^{\prime}_{ii}$,
$
H^{\prime}_{i,N+1}$, $H^{\prime}_{N+i,i}$ and $H^{\prime}_{N+i,N+i}$
of the above matrix
is
independent of the other matrix elements. The matrix composed of these four matrix elements
is
diagonalized using a unitary matrix
\begin{eqnarray}
U_{pq}=
\left\{\begin{array}{c}
u_{pq} \hspace{4ex}\mbox{for $p, q=i, N+i$} \\
\mbox{ } \\
\hspace{-4ex} \delta_{pq} \hspace{4ex}\mbox{for the others} 
\end{array}\right.,
\end{eqnarray}
where $u_{pq}$ are the four elements of the (two-dimensional) unitary matrix. The determinant to yield the eigenenergies is
\begin{eqnarray}
(E-H^{\prime}_{ii})
(E-H^{\prime}_{N+i,N+i})-H^{\prime}_{i,N+i}H^{\prime}_{N+i,i}
=0.
\end{eqnarray}
We
now
consider
physically meaningful phenomena, whose 
eigenenergies are
those
for the row with $q_{r}=1$ (the lowest oscillation case) of the aforementioned matrix
$R$,
and drop the
$q_{r}$-dependent
term from the Hamiltonian owing to small
$|\sin(\pm 2\pi q_{r} /(2N+1))|$. 
(The contribution of the string tension term
$\sigma i\Delta$ 
to the energy Hamiltonian is large in the case of not small $i$ within the strongly bound confinement regime, while $\kappa/(i\Delta)$
exhibits a
singularity for
low
$i$.) 
Additionally, we neglect the
masses
of the composite fundamental fermions,
for the cases in which the masses are small compared to the rotational energy.
We then get
the
eigenenergies
\begin{eqnarray}
\label{LESM} 
(E
+\frac{\alpha}{i\Delta}
-\sigma i \Delta)^{2}-(\frac{|\kappa|}{i\Delta })^{2}=0,
\end{eqnarray}
which states that
\begin{eqnarray}
\label{EDET} 
E=-\frac{\alpha}{i\Delta}+\sigma i \Delta + \frac{|\kappa|}{i\Delta }.
\end{eqnarray}
The above 
Coulomb term $-\alpha/(i\Delta)$ is disregarded for the extremely small $\Delta$ (in some sense beyond the regime of usual computer simulations with the larger $\Delta$) due to the asymptotic freedom of the non-Abelian field, yielding
\begin{eqnarray}
 \label{EqE} 
E= \sigma i \Delta + \frac{|\kappa|}{i\Delta },
\end{eqnarray}
 ($i$ is not the complex number but an integer of the lattice index).
\par
Here, we consider a function of the number $x$ in the real continuum in the region $x > 0$, defined by
\begin{eqnarray}
E_{x}=\frac{|k|}{x}+\sigma x=(\sigma x+\frac{|k|}{x}).
\end{eqnarray}
The function
$E_{x}$
above takes the minimum at\\
$x_{\rm m}=(|\kappa|/\sigma)^{1/2}$
as
\begin{eqnarray}
\nonumber
E_{x}^{\rm min}=\frac{|k|}{(|\kappa|/\sigma)^{1/2}}+\sigma(|\kappa|/\sigma)^{1/2}
\end{eqnarray}
\begin{eqnarray}
=2(|\kappa|\sigma)^{1/2}.
\end{eqnarray}
This $x_{\rm m}$ is (considering $N\Delta > x_{\rm m}$) measured with the lattice spacing $\Delta$ as
\begin{eqnarray}
x_{\rm m}=i_{\rm m}\Delta+\epsilon_{\rm E},
\end{eqnarray}
where $i_{\rm m}$ is an integer and $\epsilon_{\rm E}$ corresponds to a residual denoted as
\begin{eqnarray}
-\frac{\Delta}{2} \leq \epsilon_{\rm E} < \frac{\Delta}{2}.
\end{eqnarray}
In the limit as $\Delta \rightarrow 0$, the residual $\epsilon_{\rm E}$ vanishes ($\epsilon_{\rm E} \rightarrow 0$), and the eigenenergy
$E$
in Eq. (\ref{EqE})
at $i=i_{m}$ approaches the minimum value $E_{\rm min}$ equal to $E_{x}^{\rm min}$, giving
\begin{eqnarray}
 \label{ESq}
E_{\rm min}^{2}
=(E_{x}^{\rm min})^2
=4 |\kappa|\sigma,
\end{eqnarray}
(which is
independent of $\Delta$).
\par
We note that $|\kappa| $ is the absolute value of the integer-type relativistic quantum number,
which includes a sign and originates 
from the Dirac equation.
It is noteworthy that
only the secular equation gives rise to the absolute value.
\par
We compare our equality
Eq. (\ref{ESq})
given above with that obtained from another theoretical method \cite{Miya, Maki} using the 
classical mechanical
Hamiltonian in spherical coordinates, which is briefly summarized as follows.
The relativistic
classical mechanical
Hamiltonian comprising the kinetic energy and a linear potential is denoted as
\begin{eqnarray}
\label{EMFM} 
H^{(\rm
cl)
}=(P^2+m^2)^{1/2}+\sigma r,
\end{eqnarray}
where $P$ is the relativistic momentum of a fundamental particle. For
small mass
compared to the rotational energy,
the
classical mechanical
Hamiltonian, which was described above, is reduced to
\begin{eqnarray}
H^{(\rm 
cl)
}=P+\sigma r. 
\end{eqnarray}
Using the rotational quantity $J$, which roughly corresponds to $|\kappa|$ and which is written by
\begin{eqnarray}
\label{EFMD} 
J=Pr,
\end{eqnarray}
the 
classical mechanical
Hamiltonian
$H^{(\rm
cl)
}$ 
amounts to
\begin{eqnarray}
\label{EMD}
H^{(\rm 
cl)
}
=\frac{J}{r}+\sigma r.
\end{eqnarray}
The energy minimum
$E^{(\rm 
cl)
}_{\rm min}$
of
$H^{(\rm 
cl)
}$ 
also occurs at 
\begin{eqnarray}
r=(\frac{J}{\sigma})^{1/2},
\end{eqnarray}
to give
\begin{eqnarray}
\label{EMTO} 
(E^{(\rm 
cl)
}_{\rm min})^2
=4 J\sigma.
\end{eqnarray}
The above relation essentially coincides with the aforementioned equality
Eq. (\ref{ESq}),
which is consistent with the 
principal properties of the
Regge trajectory \cite{ColReg}. 
\par
From the experimentally observed slope
\begin{eqnarray}
\frac{d
|
\kappa
|
}{d (E_{\rm min}^2)}=0.93,
\end{eqnarray}
in natural units for the Regge trajectory \cite{ColReg, Naga}, the equality $|\kappa|=E_{\rm min}^{2}/(4\sigma)$
results in
$1/(4\sigma)=0.93$ GeV$^{-2}$ to yield $\sqrt{\sigma}=518.5$ MeV. If we use the relation $\sqrt{\sigma}=2.255\Lambda_{\rm MOM}$,
derived analytically in our previous paper \cite{Fuku14}
where
$\Lambda_{\rm MOM}$ corresponds to the scale-invariant energy of quantum chromodynamics (QCD),
we arrive at an $\Lambda_{\rm MOM}$ of $229.9$ MeV. This
is larger than
the
$186$ MeV 
calculated
for
a
smaller setting
of
$\sqrt{\sigma}=420$ MeV in our previous paper \cite{Fuku14}. These values are
consistent
with the observed QCD scale-invariant energy
of
around 213 MeV \cite{MPes}.
\par

\subsection{Discussions}

Here, we add some discussions concerning the present approach and results obtained in the previous section and subsection.
Unlike our approach, the 
classical mechanical Hamiltonian
with a linear potential and repulsive rotational potential \cite{Miya,Maki}, described by Eqs. (\ref{EMFM})-(\ref{EMTO}) in this paper,
has no basis in the Dirac/Yang-Mills equations. Furthermore, concerning the Regge trajectory expressed as a function of the rotational quantum number for eigenenergies (masses) of the pair of the constituent fundamental fermion and antifermion, the mechanism yielding large binding energies compared with the composite fermion masses has not yet been
fully
clarified by other theoretical/numerical approaches. 
\par
In contrast, we used the Dirac equation in spherical coordinates at the first quantization level, considering the following points. First, because spherical coordinates differ from orthogonal coordinates, the eigenenergies are derived as a function of the relativistic quantum number, considering the expression of the Regge trajectory. Second, the procedure for deriving the solution of the Dirac equation shows that the mass (binding eigenenergy) of the pair of the fundamental fermion and antifermion has its origin in the linear force between the particles and the angular potential proportional to the relativistic angular quantum number. Therefore, the 
eigenequation in the form of determinant gives rise to the
lowest
energy 
as a function of the relativistic quantum number
corresponding to the
classical mechanical Hamiltonian energy for the principal properties of the Regge trajectory
\cite{Miya,Maki}, which is presented in the part containing Eqs. (\ref{EMFM})-(\ref{EMTO}) in this article.
\par
We note that our formalism for the Dirac equation in the present paper uses the linear potential with the attractive Coulomb potential, which was directly derived from the Yang-Mills equation using the path integral at the second quantized field-theoretic level.
The potential used here was
calculated non-perturbatively using the Wilson loop, which has all orders of boson contributions. The Wilson loop $W_{\rm Q}$ gave rise to (Coulomb potential + linear potential) in Eq. (\ref{DPOT}), as follows:
\begin{eqnarray}
V_{\rm W}(r)=-\frac{\ln [W_{\rm Q}(r)]}{t_{2}-t_{
1
}}, 
\end{eqnarray}
where
$t_{2} - t_{1}$ is the time interval used in calculating the Wilson loop.
If we use the Wilson loop derived in our previous paper \cite{Fuku14}, the linear force part, which was provided analytically, is expressed as
\begin{eqnarray}
W_{\rm QL}(r)=
\exp[-\sigma r (t_{2}-t_{1})],
\end{eqnarray}
where $t_{1}$ is a small quantity.
Then, the detailed form of Eq. (\ref{EDET}),
when the rotational energy is large
compared to the fermion mass, becomes
\begin{eqnarray}
E=V_{\rm W}(i\Delta)+\frac{|\kappa|}{i\Delta}.
\end{eqnarray}
The eigenenergies of the Dirac equation obtained in this paper can be larger than the
masses of the composite fermion and antifermion, 
when the masses are small compared to the rotational energy.
The Polyakov line shows the deconfinement at high temperatures, and if we use the result of
our previous paper \cite{Fuku14}, the Polyakov line $P_{\tau}$
which we analytically derived is expressed as
\begin{eqnarray}
\nonumber
P_{\tau}=\cos\{\arccos[\exp(-\sigma r\tau)]
 -\arccos[\exp(-\sigma r\tau_{\epsilon})]\}
\end{eqnarray}
\begin{eqnarray}
\approx \cos\{\arccos[\exp(-\sigma r\tau)]\},
\end{eqnarray}
with $\tau =1/(k_{B}T)$ ($k_{\rm B}$ and $T$ are the Boltzmann constant and temperature, respectively) and $\tau_{\epsilon}$ being a small quantity. Then,
\begin{eqnarray}
\label{EPOL}
\epsilon_{\rm q}=-\ln(P_{\tau}) \approx -\ln[\exp(\frac{-E_{\rm B}}{k_{\rm B}T})]=\frac{E_{\rm B}}{k_{\rm B}T}
=\frac{\sigma r}{k_{\rm B}T},
\end{eqnarray}
where $\epsilon_{\rm q}$ is the binding energy of the pair of the fundamental fermion and antifermion, and
$E_{\rm B}=\sigma r$.
Equation ($\ref{EPOL}$)
shows that $\epsilon_{\rm q}$ is small at high temperatures
and the deconfinement of paired fermions occurs in some sense. The
classical mechanical Hamiltonian describes the system at absolute zero (temperature) and did not treat
this deconfinement
at high temperatures.
\par
Before the next discussions, we briefly refer to the quenched case, which takes into account the Okubo-Zweig-Iizuka (OZI) rule [29-31], implying that the further fermion-antifermion pair creation is suppressed.
(The quenched case corresponds to the following approximation in the case of the path integral with respect to fermion Grassmann numbers $\bar{\Psi}$ and $\Psi$, including a matrix $M_{\rm f}$,
\begin{eqnarray}
\int d \bar{\Psi} d\Psi \exp(- \bar{\Psi}M_{\rm f}\Psi)
={\rm det}(M_{\rm f}),
\end{eqnarray}
which is set to unity.)
\par
In the operator formalism at Euclidian time $t$, the Green's (two-point correlation) function is given by
\begin{eqnarray}
\label{GreF}
G_{t0}(t)=<0|\hat{\cal{H}}_{\rm B}(t)
{\hat{\cal{H}}}^{\dag}_{\rm B}(0)|0>,
\end{eqnarray}
where $<0|$ and $|0>$ is the ground state vacuum and $\hat{\cal{H}}_{\rm B}$ is the Heisenberg-type Hamiltonian operator of the bound state, which is expressed in terms of the energy operator $\hat{E}$ as
\begin{eqnarray}
\hat{\cal{H}}_{\rm B}(t)=\exp(\hat{E}t)\hat{\cal{H}}_{\rm B}(0)\exp(-\hat{E}t).
\end{eqnarray}
Using Eqs. (\ref{SolF}) and (\ref{SolG}), the $s$-th solution of the Dirac equation $|s>$ with eigenenergy $E_{(s)}$ is denoted as
\begin{eqnarray}
|s> \propto
\left[\begin{array}{c}
\frac{G_{(s)}}{r} \\
            \\
\frac{F_{(s)}}{r}
\end{array}\right].
\end{eqnarray}
(The above radial functions are multiplied by spin-angular components and a constant factor for the center of mass $\exp(iP_{\mu}X_{\mu})$, with $X_{\mu}$ and $P_{\mu}$ being the position coordinates for the center of mass and its momentum, respectively.) For the quenched case, the Green's function given by Eq. (\ref{GreF})
yields,
using the above solution $|s>$,
\begin{eqnarray}
\nonumber
G_{t0}(t)=\sum_s <0|\hat{\cal{H}}_{\rm B}(t)|s><s|
{\hat{\cal{H}}}^{\dag}_{\rm B}(0)|0>
\end{eqnarray}
\begin{eqnarray}
=\sum_s|<0|\hat{\cal{H}}_{\rm B}(0)|s>|^2
\exp(-E_{(s)} t),
\end{eqnarray}
where the eigenenergy of the Dirac equation has appeared as a decay constant for the Euclidian time.
\par
Thus, the solutions of the Dirac equation associated with the eigenenergies enter into the operator formalism with the help of the OZI rule (the suppression of the further fermion-pair creation). As previously mentioned, the Dirac equation indicates that the
lowest
mass 
(expressed as an eigenenergy as a function of the relativistic quantum number when the rotational energy is larger than the constituent particle masses)
of the pair of the fundamental fermion and antifermion originates in the confining linear potential and the angular potential proportional to the relativistic quantum number. The mass of the pair is allowed to be larger than the masses of the composite particles. Furthermore, the deconfinement feature at high
temperatures
in some sense is described by the Polyakov line according to the Yang-Mills theory, as presented in our previous paper \cite{Fuku14}.
\par
%
%

\section{Conclusions}

We
have presented a formalism for the Dirac field under a confining linear potential using basis functions localized in
the
spacetime
continuum,
which formulates fields of finite degrees of freedom.
A
given linear potential is that from the Wilson loop analysis for
a non-Abelian Yang-Mills field. The Hamiltonian matrix
has been
analytically diagonalized
with the use of two sequential unitary transformations,
thus
yielding the eigenenergies of the confined fundamental fermion-antifermion pair.
The lowest eigenenergy (as a function of the relativistic quantum number for the large rotation energy compared to the composite particle masses) is
proportional to the string tension and the Dirac's relativistic quantum number
related
to the total angular momentum, which is consistent with the expectation.

\par

\end{multicols}


\begin{thebibliography}{}

\bibitem{Fuku84}
K.~Fukushima, Phys. Rev. D {\bf30}, 1251 (1984).
\bibitem{Fuku14}
K.~Fukushima and H.~Sato, Bulg. J. Phys. {\bf 41}, 142 (2014); \\
arXiv:1402.0450 (arXiv:1402.0450v5). \\
Freely available at \\
http://www.bjp-bg.com/papers/ \\
bjp2014\_2\_142-171.pdf
\bibitem{BenMS}
C.~M.~Bender, K.~A.~Milton and D.~H.~Sharp, Phys.\ Rev.\ Lett. {\bf 51}, 1815 (1983).
\bibitem{PRam}
P.~Ramond, {\it Field Theory: A Modern Primer}, 3rd prn. (Benjamin, MA, 1982).
\bibitem{Aber}
E.~S.~Aber and B.~W.~Lee, Phys. Rep. {\bf 9C}, 1 (1973).
\bibitem{MPes}
M.~E.~Peskin and D.~V.~Schroeder, {\it An Introduction to Quantum Field Theory} (Addison-Wesley, Reading, MA, 1995).
\bibitem{Act79}
A.~Actor, Rev. Mod. Phys. {\bf 51}, 461 (1979).
\bibitem{Ref3}
K.~G.~Wilson, Phys. Rev. D {\bf 10}, 2445 (1974).
\bibitem{Ref4}
K.~G.~Wilson and J.~B.~Kogut, Phys. Rep. {\bf 12C}, 75 (1974).
\bibitem{Ref5}
J.~B.~Kogut, Rev. Mod. Phys. {\bf 51}, 659 (1979).
\bibitem{Ref6}
J.~B.~Kogut, Rev. Mod. Phys. {\bf 55}, 775 (1983).
\bibitem{Ref7}
M.~Creutz, Phys. Rev. D {\bf 21}, 2308 (1980).
\bibitem{Ref8}
M.~Creutz, L.~Jacobs and C.~Rebbi, Phys. Rep. {\bf 95}, 201 (1983).
\bibitem{Ref9}
J.~-M.~Drouffe and C.~Itzykson, Phys. Rep. {\bf C38}, 133 (1978).
\bibitem{Ref10}
J.~-M.~Drouffe and J.~-B.~Zuber, Phys. Rep. {\bf 102}, 1 (1983).
\bibitem{HRot}
H.~J.~Rothe, {\it Lattice Gauge Theories: An Introduction}, 3rd edition (World Scientific Publishing, Singapore, 2005).
\bibitem{HaSca}
A.~Hasenfratz and P.~Hasenfratz, Phys. Lett. B {\bf 93}, 165 (1980).
\bibitem{Polya}
A.~M.~Polyakov, Phys. Lett. {\bf 72B}, 477 (1978).
\bibitem{Kabu}
M.~Kaburagi, M.~Kawaguchi, T.~Morii, T.~Kitazoe and J.~Morishita, Z. Phys. C {\bf 9}, 213 (1981).
\bibitem{AbeF}
S.~Abe and T.~Fujita, Nucl. Phys. {\bf A475}, 657 (1987).
\bibitem{Tez91}
H.~Tezuka, J. Phys. A {\bf 24}, 5267 (1991).
\bibitem{Tez13}
H.~Tezuka, AIP Advances {\bf 3}, 082135 (2013).
\bibitem{ColReg}
P.~D.~Collins and E. J. Squires,
{\it Regge Poles in Particle Physics},
Springer Tracts in Modern Physics, Vol. 45
(Springer-Verlag, Berlin, 1968).
\bibitem{Miya}
H.~Miyazawa, {\it Quark
confinement
}
(in Japanese), in Buturigaku Saizensen (Physics Frontier), Vol. 1, ed.\\
Y.~Ohtsuki, (Kyoritsu Shuppan, Tokyo, 1982) p. 1.
\bibitem{Maki}
Z.~Maki and K.~Hayashi, {\it Soryushi Butsurigaku (Elementary Particle Physics)} (in Japanese), (Maruzen, Tokyo, 1995).
\bibitem{Schf}
L.~I.~Schiff, {\it Quantum Mechanics}, 3rd edition (Mcgraw-Hill, New York,
1968).
\bibitem{Dira}
P.~A.~M.~Dirac, {\it The Principles of Quantum Mechanics}, 4th edition (Oxford University Press, Oxford, 1958).
\bibitem{Naga}
Y. Nagashima, {\it Soryusi Hyojun Riron to Jikken-teki Kiso (Standard Theory of Elementary Particles and Experimental Basis)} (in Japanese), (Asakura Publishing, Tokyo,
1999).
\bibitem{Okub}
S. Okubo, Phys. Lett. {\bf 5},
165 
%
(1963).
\bibitem{Zwei}
G. Zweig, CERN Report, No.8419/TH412 (1964).
\bibitem{Iizu}
J. Iizuka, Prog. Theor. Phys. Suppl.
{\bf 37-38}, 21 (1966).

\end{thebibliography}
\end{document}